%% file: paper.tex
\documentclass[conference,onecolumn,12pt]{IEEEtran}

\usepackage{algorithm}
\usepackage[noend]{algpseudocode}
\usepackage{amsfonts}
\usepackage{amsmath}
\usepackage{amssymb}
\usepackage{amstext}
\usepackage{amsthm}
\usepackage{array}
\usepackage{bigdelim}
\usepackage{booktabs}
\usepackage{caption}
\usepackage{footmisc}
\usepackage{graphicx}
\usepackage[hidelinks]{hyperref}
\usepackage{lipsum}
\usepackage{listings}
\usepackage{mathpartir}
\usepackage{mathtools}
\usepackage{multirow}
\usepackage[numbers, sort]{natbib}
\usepackage{pifont}
\usepackage{setspace}
\usepackage{stmaryrd}
\usepackage{subcaption}
\usepackage{tikz}
\usepackage{url}
\usepackage{xcolor}
\usepackage{xspace}

\renewcommand{\paragraph}[1]{\noindent\textbf{#1}}

\theoremstyle{definition}

\algnewcommand{\Input}[1]{\Statex \textbf{Input} #1}
\algnewcommand{\Modifies}[1]{\Statex \textbf{Modifies} #1}
\algnewcommand{\Returns}[1]{\Statex \textbf{Returns} #1}

\newcommand{\cmark}{\ding{51}}
\newcommand{\xmark}{\ding{55}}
\newcommand{\tcmark}{$^\text{\cmark}$}
\newcommand{\txmark}{$^\text{\xmark}$}

\begin{document}

\title{FastFlip: Compositional Error Injection Analysis}

\author{
\IEEEauthorblockN{Keyur Joshi, Rahul Singh, Tommaso Bassetto,\\Sarita Adve, Darko Marinov, and Sasa Misailovic}
\IEEEauthorblockA{University of Illinois Urbana-Champaign}
}

\maketitle

\begin{abstract}
Instruction-level error injection analyses aim to find instructions where errors often lead to unacceptable outcomes like Silent Data Corruptions (SDCs).
These analyses require significant time, which is especially problematic if developers wish to regularly analyze software that evolves over time.

We present FastFlip, a combination of empirical error injection and symbolic SDC propagation analyses that enables fast, compositional error injection analysis of evolving programs.
FastFlip calculates how SDCs propagate across program sections and correctly accounts for unexpected side effects that can occur due to errors.
Using FastFlip, we analyze five benchmarks, plus two modified versions of each benchmark.
FastFlip speeds up the analysis of incrementally modified programs by $3.2\times$ (geomean).
FastFlip selects a set of instructions to protect against SDCs that minimizes the
runtime cost of protection while protecting against a developer-specified target fraction of
all SDC-causing errors.
\end{abstract}


\input{fastflip}

\section*{Acknowledgements}

The research presented in this paper was funded through NSF Awards CCF-1846354
and CCF-1956374. We would also like to thank Dr. Abdulrahman Mahmoud for his
help with Approxilyzer.

\bibliographystyle{IEEEtranS}
\bibliography{references}

\end{document}

%% file: fastflip.tex
\input{ff-defs}

\section{Introduction}

As conventional technology scaling approaches its end, hardware is becoming
increasingly susceptible to errors~\cite{Borkar2005, 10.1145/3600006.3613149}.
Unlike crashes, timeouts, or detectable output corruptions, the presence of
Silent Data Corruptions (SDCs) in program outputs caused by such hardware errors
may not become apparent until long after the program has terminated. Researchers
have proposed various hardware and software techniques to protect programs
against SDCs. \emph{Software protection techniques} such as instruction or task
duplication~\cite{swift, drift, Shoestring2010, nzdc} are particularly
attractive as they can be used on existing hardware. They typically use
additional computational resources to detect SDCs and/or recover from them. To
limit their time and energy overhead, it is necessary to selectively use such
techniques on the parts of the program that are most vulnerable to SDCs.

Finding vulnerable instructions is the task of instruction-level error injection
analyses. These analyses inject errors into different architectural components
of a simulated CPU at various points in a program's execution, and record the
effect of the error on the program output. For targeted SDC protection, the
analysis must provide information on how errors at \emph{each instruction} in
the program affect the output (e.g.,~\cite{relyzer, approxilyzer}). Such
detailed per-instruction analyses are time-consuming, requiring thousands of
core-hours even for small programs and inputs. While sampling-based tools
like~\cite{llfi, avgi} are faster, they cannot provide the necessary
vulnerability information for every instruction.

This high cost of per-instruction error injection analysis is concerning because
modern programs are continuously evolving; developers change code to fix bugs,
add features, or add optimizations manually or using compilers. The modified
program sections react differently to errors, so they must be re-analyzed.
However, current error injection analyses must be rerun on the whole program,
not just the modified sections, before applying software protection techniques.

An intuitive solution is to leverage the compositional nature of programs by
dividing them into multiple dynamic sections (such as function calls or
executions of code blocks or nested loops), applying the error injection
analysis on each section separately, and then combining the results. However,
such a compositional error injection analysis must overcome multiple challenges:
1)~SDCs in the output of one section must be propagated through downstream
sections to determine the SDCs in the final output of the program, and 2)~an
error in one section can corrupt data that will be used only by downstream
sections, thus causing unexpected side effects that do not affect the current
section's output. To the best of our knowledge, none of the previously proposed
analyses (e.g.,~\cite{DweikRAEs, MeRLiN, swatsim, approxilyzer, gem5A,
SmartInjector, failstar, gefinmafin, gemfiParasyris, flipit, llfi, hamartia,
avf_online}) can selectively analyze only the modified parts of the program.

\paragraph{Our work.}
We present FastFlip, a unique combination of empirical (error injection) and
symbolic (SDC propagation) techniques, which makes it the first approach for
compositional and incremental error injection analysis of evolving programs.
Given a program and an input that FastFlip has not previously analyzed, FastFlip
proceeds as follows: First, FastFlip analyzes the sections of the program for
that input with a per-instruction error injection analysis to determine the
outcome of each possible injected error. Second, FastFlip uses a local
sensitivity analysis to determine how each section propagates and amplifies SDCs
within its input, and then uses an SDC propagation analysis to determine how
SDCs propagate across different sections to affect the final output. Third,
FastFlip combines the results of the error injection analysis and the SDC
propagation analysis to determine how injected errors in any section affect the
final output. FastFlip uses this information to select a set of static
instructions to protect, such that the dynamic cost of protecting the selected
instructions is minimized, while ensuring that the degree of protection against
SDC-causing errors is above a developer-specified threshold. FastFlip correctly
accounts for side effects that only occur due to errors.

If developers modify a program after FastFlip has analyzed it, FastFlip can
reuse large portions of its analysis results. In particular, FastFlip must only
rerun the expensive error injection analysis on the modified program section,
and any downstream sections whose input changes as a result of the modification.
Consequently, FastFlip saves significant analysis time with each program
modification.

We instantiate FastFlip using the Approxilyzer~\cite{approxilyzer}
per-instruction error injection analysis and the Chisel~\cite{chisel} SDC
propagation analysis. Approxilyzer focuses on analyzing the effects of hardware
errors that manifest within CPU architectural registers as bitflips. While
analyses exist that also inject errors in other architectural components
(e.g.,~\cite{avgi}), they rely on sampling, and would be impractical if used to
obtain per-instruction results. We analyze five benchmarks with FastFlip, each
with three or more separately analyzed sections, and compare against a baseline
Approxilyzer-only approach, which treats the entire program as a single section.

We evaluate the utility of protecting the static instructions selected by
FastFlip and Approxilyzer using two metrics. The \emph{value} of protecting a
selection of instructions is the total probability that an SDC-causing error
will be detected through said protection. The \emph{cost} of a selection is
roughly equal to the runtime overhead of protecting the selected instructions.
For our evaluation, we use the value and cost model from~\cite{HariAdve2012},
which assumes that value is proportional to the number of SDC-causing errors
that can occur in the selected instructions, and that cost is proportional to
the number of dynamic instances of the selected instructions that must be
protected. FastFlip's selection provides a protection value within 0.3\%
(geomean) of the target protection value for a cost of protection within 0.7\%
(geomean) of the cost of protecting Approxilyzer's selection. FastFlip can also
use more complex value and cost metrics to make its selection.

Next, we make two modifications to each benchmark: a \emph{small}, simple
modification and a \emph{large} modification that uses a lookup table. FastFlip
does not need to inject errors in the unmodified sections, providing a
$3.2\times$ speedup (geomean) over Approxilyzer. \emph{Crucially, FastFlip
provides this speedup with minimal additional loss in protection value or
increase in cost with respect to Approxilyzer.} We also experiment with a
modification that adds error detection mechanisms to a benchmark; FastFlip is
able to efficiently verify that such a modification significantly decreases the
likelihood of SDCs occurring due to errors in the protected sections.

\section{Background}

\subsection{Error injection analyses}
\label{ff:sec:bkg:inja}

Error injection analyses analyze the effect of injecting errors such as bitflips
in the program execution. The analysis first enumerates error injection sites in
the correct dynamic trace of the program execution, which is a sequential list
of instructions executed by a program for a particular input. Depending on the
analysis, these error sites can be bits in the source and destination registers
in each instruction, bits in control registers, caches, etc. The analysis then
injects errors at each site one at a time, and then executes the rest of the
program (which may deviate in control flow from the correct execution), to
record the effect of the error on the final output. Such analyses can operate at
different levels of abstraction, including hardware, assembly, and IR (e.g.,
RAEs~\cite{DweikRAEs}, Approxilyzer~\cite{approxilyzer}, and
FlipIt~\cite{flipit} respectively.). An error can have five possible effects on
the output of the program:
\begin{itemize}
\item The error is \emph{masked}, i.e., the program output is unaffected.
\item The error causes a program \emph{crash}, i.e., the program terminates
unexpectedly.
\item The error greatly extends the program runtime (e.g., by creating a long
loop), causing a \emph{timeout}.
\item The error changes the program output in a \emph{detectable} manner (e.g.,
by producing a misformatted output or an output that is outside the expected
range).
\item The error changes the program output in an \emph{undetectable} manner,
i.e., it causes a \emph{Silent Data Corruption} (SDC).
\end{itemize}
For the last case, the analysis typically reports the magnitude of the output
SDC, calculated using an applicable SDC metric. The result of the analysis is a
map from each error site to the outcome of an error at that site. The crash,
timeout, and detectable error outcomes can be handled through detection and
recovery mechanisms such as checkpoints. The SDC outcome is the most dangerous;
the presence of an SDC may not be detected until long after the program has
terminated. However, many applications are capable of tolerating small SDCs up
to a threshold magnitude $\errlim$. Thus, similar to Approxilyzer, we further
categorize SDCs above this threshold as \emph{SDC-Bad} and smaller SDCs as
\emph{SDC-Good}.

Some error injection analyses use sampling to inject errors at random points in
the program (e.g.,~\cite{ashrafsc15, PapadimitriouDemystifying}). With a
sufficient number of samples, these analyses provide statistically significant
information on the relative frequencies of the above outcomes. Other error
injection analyses aim to provide information on the outcome of error injections
at \emph{all} potential error sites of a particular class within a program's
execution (e.g.,~\cite{relyzer,approxilyzer}). These non-sampling analyses are
slower, but FastFlip can use their detailed results to find optimal static
instructions to protect against SDCs.

\subsection{SDC propagation analyses}
\label{ff:sec:bkg:propanalysis}

SDC propagation analyses propagate SDCs present in a program's input, or SDCs
that occur during program execution, to calculate their effect on the program's
final output. These can be either \emph{forward} analyses, or \emph{backward}
analyses (e.g., Diamont~\cite{diamont} and Chisel~\cite{chisel} respectively),
which propagate SDC bounds in the respective directions through the program. An
SDC bound $\Delta(\anoutp) \le f(\Delta(\aninp))$ states that the SDC in the
output $\anoutp$ of a section of code, calculated by an appropriate SDC metric
$\Delta$, is bounded by a function $f$ of the SDC in the input $\aninp$.

\paragraph{Sensitivity analysis.}
Sensitivity analysis~\cite{sensanalysis} is a component of SDC propagation
analyses that is used to determine how a section of code amplifies SDCs in its
input. In particular, \emph{local} sensitivity analysis focuses on determining
the effect of perturbations around a single input value.

A local sensitivity analysis varies an input $x_0$ to a program section
$\trcsec$ by various amounts $\errvar$ up to some maximum perturbation
$\errvar_{\max{}}$. The analysis executes $\trcsec$ to calculate the output
perturbation and divides it by the input perturbation to calculate the SDC
amplification factor $K$, which is the Lipschitz constant for $\trcsec$ at
$x_0$:
\begin{equation*}
K = \max\limits_{\errvar\le\errvar_{\max{}}} \frac{\left|\trcsec(x_0+\errvar)-\trcsec(x_0)\right|}{\errvar}
\end{equation*}
If the program is differentiable, it may be possible to analytically calculate
$K$ via static analysis (e.g.,~\cite{chaudhuri2011proving, deepj}). In the
general case, we can approximate $K$ by sampling a set of $\errvar$ values.
Increasing the number of samples increases the quality of the approximation of
$K$, though the rate of convergence also decreases, leading to diminishing
returns.

\subsection{Protecting against SDCs in software}
\label{ff:sec:bkg:sdcprot}

Software techniques for protecting against SDCs have the advantage that they can
be used on existing, non-specialized hardware. While systems can detect crashes
and detectable errors (e.g., incorrectly formatted outputs) with relatively
lightweight mechanisms, SDCs are harder to detect by nature. Detecting SDCs
typically involves re-executing parts of the program and comparing the outcomes.
Coarse grained solutions re-execute entire tasks~\cite{raft, 641939, aloe},
while fine grained solutions re-execute individual instructions or small blocks
of instructions~\cite{swift, drift, Shoestring2010, nzdc}.

Duplicating tasks or instructions does not necessarily double the program's
resource usage. SWIFT~\cite{swift} shows how instruction reordering by the
compiler and the CPU can reduce the runtime cost of instruction duplication to
an average of 41\% of the program's original runtime. DRIFT~\cite{drift} checks
the results of multiple duplicated instructions at once to further increase
instruction level parallelism and reduce the runtime overhead to 29\% on
average.

We can also choose to selectively protect only the instructions that are most
vulnerable to SDCs. We can use error injection analyses that inject errors in
all instructions~\cite{relyzer,approxilyzer} to find such vulnerable
instructions and selectively protect them. Each instruction has a value/cost
tradeoff associated with it: the cost of protecting a static instruction using
instruction duplication is roughly proportional to the number of dynamic
instances of the instruction that will have to be duplicated, while the value is
roughly proportional to the likelihood that an error that causes an SDC will
occur in that instruction.

\section{Example}
\label{ff:sec:example}

Lower-Upper decomposition (LU) is a key matrix operation that is used to solve
systems of linear equations and to compute the matrix inverse or determinant.
The blocked LU decomposition algorithm improves performance by dividing the
matrix into blocks and processing a small fraction of blocks at a time. The
algorithm consists of an outer loop whose loop body has four logical sections:
\begin{enumerate}
\item The algorithm decomposes a block $B$ on the matrix diagonal
\item The algorithm updates blocks below $B$ in the matrix
\item The algorithm updates blocks to the right of $B$ in the matrix
\item The algorithm updates blocks below \emph{and} to the right of $B$
\end{enumerate}

Given the widespread use of LU decomposition, it is inevitable that hardware
errors will occasionally occur in large computations that use this operation.
While memory can be protected using ECC, data in registers is more vulnerable.
If a bitflip error causes a detectable effect, then the software can be
restarted or a previous checkpoint can be reloaded. However, if a bitflip causes
an SDC, the corruption may not be detected until much later. Thus, we wish to
protect the program against SDCs using techniques such as instruction
duplication~\cite{swift, drift, Shoestring2010}.

We can use an error injection analysis like Approxilyzer~\cite{approxilyzer,
gem5A} to analyze the effect of bitflips on the full program, and use the
results to guide protection against SDCs. However, if developers modify one
section of the program during development (e.g., to add optimizations), then we
must rerun the analysis on the full modified program, which requires thousands
of core-hours even for simple programs. We cannot naively re-analyze only the
modified section with Approxilyzer, because the modified section most likely
responds differently to injected errors, which leads to a different distribution
of injection outcomes at the end of the full program.

To solve this issue, we present FastFlip, a compositional approach for error
injection analysis of programs. FastFlip first separately analyzes each section
of the program using Approxilyzer (to determine the effects of errors occurring
within that section) and a local sensitivity analysis (to determine how the
section propagates existing SDCs). FastFlip then combines the analysis results
for each section using Chisel~\cite{chisel}, an SDC propagation analysis, to
calculate the end-to-end SDC characteristics of the program. Lastly, FastFlip
uses these results to find optimal instructions to protect against SDCs.

We demonstrate the FastFlip approach on the blocked LU decomposition
implementation from the Splash-3 benchmark suite~\cite{splash3} for a sample $16
\times 16$ input matrix with an $8 \times 8$ block size.
Minotaur~\cite{minotaur} has shown this input configuration to be sufficient for
100\% program counter coverage. We assume that a single bitflip occurs during
the program execution at an error site chosen uniformly at random from the
program's correct dynamic trace within an architectural register.

\paragraph{Per-section analysis.}
FastFlip first runs Approxilyzer on each section $\trcsec$ of the full program
execution $\fulltrc$. Approxilyzer determines the effect of injecting a bitflip
into each bit in each register in each dynamic instruction in $\trcsec$ on the
output of $\trcsec$, and stores these outcomes for use in later steps. For SDC
outcomes, FastFlip also stores the SDC magnitude, which is the maximum absolute
error among all output elements of $\trcsec$.

Besides the SDC caused by bitflips, sections can also propagate and amplify SDCs
caused by bitflips in previous sections. For example, FastFlip calculates that
if the input to the first section in the second iteration of the outermost loop
($s_{2,1}$) has a pre-existing SDC of magnitude $\Delta(i_{s_{2,1}})$, and no
bitflips occur within $s_{2,1}$, then the SDC in the output of $s_{2,1}$ will be at most
$3.2\Delta(i_{s_{2,1}})$.

Under the single bitflip error model, a section $\trcsec$ may either propagate
an existing SDC from a previous section, or the bitflip may occur within
$\trcsec$, but not both. Therefore, we can write the \emph{total SDC
specification} for the output of $\trcsec$ as the sum of the propagated SDC and
the SDC due to a bitflip within $\trcsec$. For example, using the results from
Approxilyzer and the sensitivity analysis, FastFlip calculates the following
upper bound on the effective SDC in the output of $s_{2,1}$
($\Delta(o_{s_{2,1}})$) as a function of the SDC in the input of $s_{2,1}$
($\Delta(i_{s_{2,1}})$) and the SDC potentially introduced by a bitflip in
$s_{2,1}$ ($\errvar_{s_{2,1}}$):
\begin{equation*}
\Delta(o_{s_{2,1}}) \le 3.2\Delta(i_{s_{2,1}}) + \errvar_{s_{2,1}}
\end{equation*}

\paragraph{Calculating end-to-end SDC specifications.}
FastFlip provides the total SDC specifications for all sections to Chisel, along
with a specification of how data flows between sections. Using this information,
Chisel calculates the \emph{end-to-end SDC propagation specification} for the
full LU decomposition computation:
\begin{equation*}
  \Delta(o_\textit{fin}) \le 4174.8\errvar_{s_{1,1}} + 434.3\errvar_{s_{1,2}}
  + 28.8\errvar_{s_{1,3}} + 3.2\errvar_{s_{1,4}} + \errvar_{s_{2,1}}
  + \errvar_{s_{2,2}} + \errvar_{s_{2,3}} + \errvar_{s_{2,4}}
\end{equation*}
where $\errvar_{s_{x,y}}$ represents the SDC potentially introduced into the
output of section $y$ in iteration $x$ by a bitflip in that section ($s_{x,y}$).
The coefficient of each $\errvar_{s_{x,y}}$ represents the \emph{total}
amplification of an SDC introduced by a bitflip in $s_{x,y}$ by sections
downstream of $s_{x,y}$. Under the single bitflip error model, only one of the
$\errvar_{s_{x,y}}$ variables can be nonzero at a time. FastFlip uses this
end-to-end SDC propagation specification to propagate SDCs caused by bitflips in
each section up to the final output.

\paragraph{Selecting instructions to protect.}
FastFlip adapts the value and cost model from~\cite{HariAdve2012} to select a
set of optimal instructions to protect against SDCs. FastFlip associates each
static instruction $\statinst$ in the program with a \emph{value} of protecting
it; this is the number of SDC-causing bitflips that can occur at $\statinst$.
Similarly, FastFlip associates each static instruction $\statinst$ with a
\emph{cost} of protection; this is the number of dynamic instances of
$\statinst$ in the program execution. We assume that the value and cost of
protecting a set of instructions is the sum of the value and cost of protecting
each instruction in the set. Given a target total SDC protection value, FastFlip
aims to select a subset of instructions that meet this target while also
minimizing the total protection cost. This is a \mbox{0\texttt{-}1} knapsack
optimization problem, which FastFlip solves via the standard dynamic programming
approach.

\paragraph{Comparison and target adjustment.}
We compare FastFlip's results to those of a baseline Approxilyzer-only approach.
This baseline approach performs an error injection analysis of the whole program
at once and uses the results to determine which instructions to protect. Using
the results of the baseline analysis, we can calculate FastFlip's
\emph{achieved} value, which is the value of protecting FastFlip's selection of
instructions according to the error injection outcomes of the baseline analysis.
If FastFlip's achieved value is below the target value, FastFlip adjusts the
target upwards, so that its selection of instructions to protect for this
adjusted target will successfully achieve the original target value.

\subsection{Results}

\begin{figure}
  \centering
  \includegraphics[width=\textwidth]{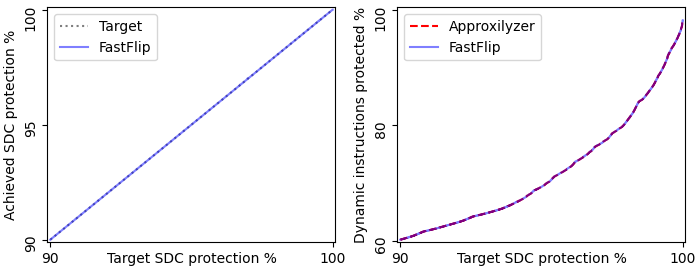}
  \caption{Protection value (left) and cost (right) comparison. Note the strong
  overlap within each plot.}
  \label{ff:fig:ex:results}
\end{figure}

\paragraph{Value.}
The left plot in Figure~\ref{ff:fig:ex:results} shows the value of protecting
FastFlip's selection of instructions against SDCs. The X-Axis shows the target
value over the range $[90\%,100\%]$. The Y-Axis shows the achieved value. The
solid blue line shows FastFlip's achieved protection value (after target
adjustment), which overlaps the dotted black line showing the target value.
FastFlip successfully achieves (or slightly overshoots) the target protection
value for the entire range of targets. Even without target adjustment, FastFlip
undershoots the target by less than 0.1\%.

\paragraph{Cost.}
The right plot in Figure~\ref{ff:fig:ex:results} compares the cost of protecting
FastFlip's selection of instructions against the cost of protecting the
Approxilyzer baseline's selection. The red dashed line and solid blue line show
the cost using Approxilyzer and FastFlip's results, respectively; the two lines
visually overlap. The maximum excess of cost of FastFlip over Approxilyzer is
below 0.1\%, even after target adjustment.

\paragraph{Modifications.}
FastFlip enables \emph{compositional analysis} by splitting the full program
Approxilyzer analysis into multiple sections. We demonstrate the benefits of
compositional analysis by performing both analyses on two modified versions of
this program. The \emph{small} modification uses a specialized version of
section 4 of the program which reduces the number of bounds checks when the
matrix size is a multiple of the block size (as is the case for our input). The
\emph{large} modification replaces the first section with a lookup table. Unlike
the Approxilyzer-only approach, which must inject errors in the full execution
of the modified program, FastFlip only needs to inject errors in the modified
section of the program, saving considerable time. FastFlip also reuses the
adjusted targets that it found for the original version of the program. FastFlip
continues to achieve the original target values with these adjusted targets, and
the excess of cost of FastFlip over Approxilyzer stays below 0.3\%.

\paragraph{Analysis time.}
FastFlip requires 694 core-hours to analyze the original version of the program,
compared to 602 core-hours for Approxilyzer. This slowdown is due to
Approxilyzer's ability to prune injections across multiple sections of the
computation by forming large equivalence classes. FastFlip cannot prune
injections to a similar extent, as it analyzes each section independently.
However, FastFlip saves a significant amount of time when subsequently analyzing
the modified versions of the program. For analyzing the program with the
\emph{small} modification, FastFlip requires 80 core-hours as opposed to 625
core-hours for Approxilyzer. Similarly, for analyzing the program with the
\emph{large} modification, FastFlip requires 94 core-hours as opposed to 441
core-hours for Approxilyzer. This shows that FastFlip is useful when analyzing
programs that evolve over time, because it saves analysis time with each
modification.

FastFlip enables efficient target adjustment by simultaneously running the
Approxilyzer baseline analysis while performing its own analysis. For the
original version of the program, the overhead of this simultaneous approach is
less than 1\% of the FastFlip-only analysis time. Since FastFlip reuses the
adjusted targets for the modified versions, it does not need the simultaneous
approach for those versions.

\section{The FastFlip approach}

\begin{figure}
\centering
\includegraphics[width=\textwidth]{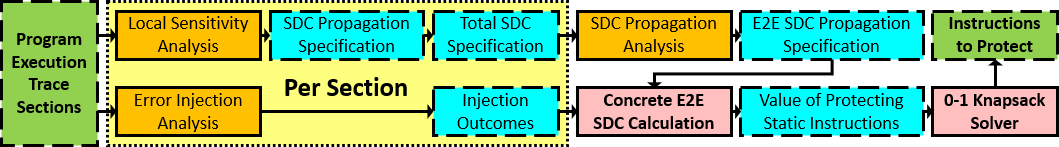}
\caption{The FastFlip approach. Green boxes with dashed outlines and bold text
are initial inputs and final results. Blue boxes with dashed outlines and normal
text are intermediate results. Pink boxes with solid outlines and bold text are
steps executed by FastFlip. Orange boxes with solid outlines and normal text are
steps executed by the sub-analyses. The dotted box shows the portion of the
approach that is applied to each section.}
\label{ff:fig:blockdiag}
\end{figure}

Figure~\ref{ff:fig:blockdiag} visualizes the FastFlip approach. First, FastFlip
performs two sub-analyses on each program section $\trcsec$ (a function call or
one execution of a code block or loop nest marked by the developer) in the full
program execution $\fulltrc$:
\begin{itemize}
\item FastFlip uses a supported error injection
analysis\footnote{\label{ff:fnt:app:toolsupport} We describe the characteristics
of supported error injection and SDC propagation analyses in
Section~\ref{ff:sec:app:usabletools}.} to determine the effect of each possible
error in $\trcsec$ and stores the outcome.
\item FastFlip uses a local sensitivity analysis to obtain an SDC propagation
specification for $\trcsec$, and converts it into a total SDC specification for
$\trcsec$.
\end{itemize}
Second, FastFlip runs a supported SDC propagation
analysis\footref{ff:fnt:app:toolsupport} over $\fulltrc$, using each section's
total SDC specification, to obtain the end-to-end SDC propagation specification
for $\fulltrc$. Third, FastFlip calculates concrete end-to-end SDC magnitudes to
find the probability of an SDC-Bad outcome associated with each static
instruction; this corresponds to the value of protecting said static
instruction. Finally, FastFlip selects a set of instructions to protect with SDC
detection mechanisms that minimizes the total cost of protection, while also
ensuring that the total value of the protection is above a developer-defined
threshold.

\subsection{Preliminaries}
\label{ff:sec:app:prelim}

As in previous works~\cite{relyzer, approxilyzer}, FastFlip assumes that 1)~the
program's input is SDC-free and 2)~exactly one error occurs during the execution
of the full program. This error can be a single or multi-bit corruption within
\emph{one} dynamic instruction.

\paragraph{Definitions.}
We use the following symbols:
\begin{itemize}
\item $\fulltrc$: full program execution

\item $\trcsec$: program section (typically a function call or one execution of
a code block or loop nest); $\trcsec \in \fulltrc$

\item $\injset$: set of all error injection sites in $\fulltrc$

\item $\injset_\trcsec$: set of all error injection sites in $\trcsec$;
$\injset_\trcsec \subseteq \injset$

\item $\outcome_\trcsec(\aninj)$: effect of an error injection $\aninj$ on the
outputs of $\trcsec$, as calculated by the error injection analysis

\item $\aninp_{\trcsec,0}, \ldots, \aninp_{\trcsec,m}$ and $\anoutp_{\trcsec,0},
\ldots, \anoutp_{\trcsec,n}$: inputs and outputs of $\trcsec$

\item $\aninp_{\fulltrc,0}, \ldots, \aninp_{\fulltrc,m}$ and
$\anoutp_{\fulltrc,0}, \ldots, \anoutp_{\fulltrc,n}$: inputs and outputs of
$\fulltrc$

\item $f_{\trcsec,k}, f_{\fulltrc,\lambda}, f_{\fulltrc,\lambda,\trcsec}$: SDC
propagation specifications calculated by the local sensitivity analysis, the SDC
propagation analysis, and FastFlip respectively

\item $\errvar_{\trcsec,k}, \errvar_{*,*}, \errvar_{\trcsec,*},
\errvar_{\bar{\trcsec},*}$: symbolic variables (or sets thereof) for SDCs
introduced into section outputs by errors

\item $p(\aninj)$: probability that the error occurs at $\aninj \in \injset$

\item $\pcmap(\aninj)$: maps $\aninj \in \injset$ to the corresponding static
instruction

\item $\errlim_\lambda$: maximum acceptable SDC for output
$\anoutp_{\fulltrc,\lambda}$ of $\fulltrc$

\item $v(\statinst)$: value of protecting static instruction at $\statinst$

\item $c(\statinst)$: cost of protecting static instruction at $\statinst$

\item $\protpcs$: set of static instructions selected for protection
\end{itemize}

\paragraph{Analysis inputs.}
FastFlip accepts the full program $\fulltrc$, its partition into sections
$\trcsec$, a specification of how data flows between sections, the probabilities
$p(\aninj)$, SDC limits $\errlim_\lambda$, and protection cost $c(\statinst)$ as
inputs. Developers can obtain the dataflow specification using standard compiler
analysis passes. Expert developers can also input this data manually.

\subsection{Error injection analysis of program sections}
\label{ff:sec:app:probcalc}

FastFlip runs an error injection analysis on each program section $\trcsec \in
\fulltrc$ to determine the effect of injected errors on the outputs of
$\trcsec$, and stores the outcome. If an injection $\aninj$ causes a detectable
outcome, such as a crash, timeout, or any clearly out-of-bounds or misformatted
output, then the outcome $\outcome_\trcsec(\aninj) = \textit{detected}$.
Otherwise, the outcome $\outcome_\trcsec(\aninj) = \{ r_0, r_1, \ldots, r_n \}$,
where $r_k$ is the magnitude of SDC (as measured by an application-specific
metric) caused by the injection $\aninj$ in output $\anoutp_{\trcsec,k}$ of
$\trcsec$. If the injection is masked for an output $\anoutp_{\trcsec,k}$, then
$r_k = 0$.

\subsection{SDC propagation analysis of program sections}

FastFlip performs a local sensitivity analysis on each program section $\trcsec
\in \fulltrc$ to calculate how it amplifies SDCs present within its input. The
local sensitivity analysis produces an \emph{SDC propagation specification} for
$\trcsec$ of the general form:
\begin{equation*}
\bigwedge_{k=0}^n \Delta(\anoutp_{\trcsec,k}) \le f_{\trcsec,k}(\Delta(\aninp_{\trcsec,0}), \ldots, \Delta(\aninp_{\trcsec,m}))
\end{equation*}
that is, for each output $\anoutp_{\trcsec,k}$ of $\trcsec$, the specification
provides the SDC bound $\Delta(\anoutp_{\trcsec,k})$ calculated as a
function $f_{\trcsec,k}$ of the SDC bounds of the inputs of $\trcsec$.

To convert this SDC propagation specification to a total SDC specification,
FastFlip adds symbolic variables $\errvar_{\trcsec,k}$ to represent the
magnitude of SDC introduced into the output $\anoutp_{\trcsec,k}$ during the
execution of $\trcsec$ as a result of an error. Under the single error model, if
the input to $\trcsec$ already contains an SDC, then the error occurred in a
previous program section, hence $\trcsec$ cannot introduce additional SDC. Thus,
for each output of $\trcsec$, we can simply write the total SDC as the sum of
the SDC due to an error in $\trcsec$ and the SDC propagated by $\trcsec$ from
its input to its output. This is the \emph{total SDC specification} for
$\trcsec$:
\begin{equation}
\label{ff:eq:app:errsectotal}
\bigwedge_k \Delta(\anoutp_{\trcsec,k}) \le f_{\trcsec,k}(\Delta(\aninp_{\trcsec,0}), \ldots, \Delta(\aninp_{\trcsec,m})) + \errvar_{\trcsec,k}
\end{equation}

\subsection{End-to-end SDC propagation analysis}

FastFlip runs an SDC propagation analysis on the full program $\fulltrc$.
FastFlip provides the analysis with the total SDC specifications from
Equation~\ref{ff:eq:app:errsectotal} for each $\trcsec \in \fulltrc$. The
analysis also uses the developer-provided dataflow specification indicating how
outputs of one section flow into the inputs of a subsequent section. The SDC
propagation analysis uses this information to calculate the relationship between
errors that occur anywhere in $\fulltrc$ to the SDC in the final outputs of
$\fulltrc$. This is the \emph{end-to-end SDC propagation specification} for
$\fulltrc$ and has the form:
\begin{equation*}
\bigwedge\limits_{\lambda=0}^{n} \Delta(\anoutp_{\fulltrc,\lambda}) \le f_{\fulltrc,\lambda}(\Delta(\aninp_{\fulltrc,0}), \ldots, \Delta(\aninp_{\fulltrc,m}), \errvar_{*,*})
\end{equation*}
where $\errvar_{*,*}$ is the list of the symbolic SDC variables across all
sections. Because we assume, as in previous work~\cite{approxilyzer}, that the
initial inputs to the program are free of SDCs, we can simplify
$f_{\fulltrc,\lambda}$ as follows:
\begin{equation*}
\bigwedge_\lambda \Delta(\anoutp_{\fulltrc,\lambda}) \le f_{\fulltrc,\lambda}(\errvar_{*,*})
\end{equation*}
Next, we create specialized versions of $f_{\fulltrc,\lambda}$ by noting that,
under the single error model, the symbolic SDC variables for only one section
can be nonzero at a time:
\begin{equation*}
f_{\fulltrc,\lambda,\trcsec}(\errvar_{\trcsec,*}) = f_{\fulltrc,\lambda}(\errvar_{\trcsec,*}, \errvar_{\bar{\trcsec},*} = \mathbf{0})
\end{equation*}
where $\errvar_{\trcsec,*}$ is the list of the symbolic SDC variables for
section $\trcsec$ and $\errvar_{\bar{\trcsec},*}$ is the list of the symbolic
SDC variables for all other sections. Finally, we rewrite the end-to-end SDC
propagation specification as:
\begin{equation}
\label{ff:eq:app:erre2etotal}
\aninj \in \injset_\trcsec \Rightarrow \bigwedge_\lambda \Delta(\anoutp_{\fulltrc,\lambda}) \le f_{\fulltrc,\lambda,\trcsec}(\errvar_{\trcsec,*})
\end{equation}
Equation~\ref{ff:eq:app:erre2etotal} states that, if FastFlip injects an error
in section $\trcsec$, then the upper bound on the SDC in output
$\anoutp_{\fulltrc,\lambda}$ of $\fulltrc$ is
$f_{\fulltrc,\lambda,\trcsec}(\errvar_{\trcsec,*})$, a function of the magnitude
of SDC in the outputs of $\trcsec$.

\subsection{Calculating the value of protecting static instructions}

FastFlip uses the injection outcomes (Section~\ref{ff:sec:app:probcalc}) and
Equation~\ref{ff:eq:app:erre2etotal} to answer the following question: \emph{For
each static instruction identified by its program counter $\statinst$ in the
full execution $\fulltrc$, what is the total probability\footnote{This is the
probability that the error occurs within $\statinst$ \emph{and} the outcome is
SDC-Bad, as opposed to the conditional probability that the outcome is SDC-Bad
when the error occurs in $\statinst$.} of error injections that result in
SDC-Bad $(|\textit{SDC}| > \errlim_\lambda)$ for any output
$\anoutp_{\fulltrc,\lambda}$ of $\fulltrc$?} The value of protecting $\statinst$
with SDC detection mechanisms is proportional to this total probability.

\begin{algorithm}
  \begin{algorithmic}[1]
    \Input{$\fulltrc$, $\injset_\trcsec$, $\pcmap(\aninj)$, $\outcome_\trcsec(\aninj)$, $\errlim_\lambda$, $p(\aninj)$: defined in Section~\ref{ff:sec:app:prelim};
           $f_{\fulltrc,\lambda,\trcsec}$: SDC propagation specification from outputs of $\trcsec$ to output $\anoutp_{\fulltrc,\lambda}$ of $\fulltrc$}
    \Returns{$\forall \statinst.\ v(\statinst)$: value of protecting the static instruction $\statinst$}
    \State $v \gets \{\;\forall \statinst.\ \statinst \mapsto 0\;\}$
    \For{$\trcsec$ in $\fulltrc$}
      \For{$\aninj$ in $\injset_\trcsec$}
        \State $\statinst \gets \pcmap(\aninj)$
        \If{$\outcome_\trcsec(\aninj) \ne \textit{detected}$}
          \If{$\exists \lambda.\ f_{\fulltrc,\lambda,\trcsec}(\outcome_\trcsec(\aninj)) > \errlim_\lambda$}
            \State $v(\statinst) \gets v(\statinst) + p(\aninj)$
          \EndIf
        \EndIf
      \EndFor
    \EndFor
    \State $v_\textit{total} = \Sigma_\statinst v(\statinst)$
    \State $\forall \statinst.\ v(\statinst) \gets v(\statinst) / v_\textit{total}$
  \end{algorithmic}
  \caption{Finding the value of protecting a static instruction}
  \label{ff:code:app:protvalue}
\end{algorithm}

Algorithm~\ref{ff:code:app:protvalue} shows how FastFlip calculates the value
$v(\statinst)$ of protecting a static instruction $\statinst$. For each error
injection in each section, FastFlip checks if the error results in a detectable
outcome. If not, FastFlip uses Equation~\ref{ff:eq:app:erre2etotal} to calculate
upper bounds on the SDCs in the outputs of $\fulltrc$ as a function of the SDCs
in the outputs of $\trcsec$. If any SDC is SDC-Bad, FastFlip adds the
probability of that error to the value of protecting $\statinst$. Lastly,
FastFlip rescales the values so that the total value of protecting all static
instructions is equal to 1.

\subsection{Finding an optimal set of instructions to protect}
\label{ff:sec:app:optimalprot}

The value $v(\statinst)$ of protecting the static instruction $\statinst$
calculated using Algorithm~\ref{ff:code:app:protvalue} and the corresponding
protection cost $c(\statinst)$ together comprise a value and cost model similar
to the one from~\cite{HariAdve2012}. FastFlip uses $v(\statinst)$ and
$c(\statinst)$ as inputs to a \mbox{0\texttt{-}1} knapsack optimization problem.
FastFlip assumes that the value and cost of protecting a set of instructions is
the sum of the value and cost of protecting each instruction in the set. Given a
developer-defined target total protection value $\targetval$, FastFlip solves
the knapsack problem via the standard dynamic programming approach to select a
set of static instructions $\protpcs$ to protect that minimizes the total
protection cost:
\begin{equation*}
\textit{Minimize} \sum_{\statinst \in \protpcs} c(\statinst) \quad \textit{such that} \sum_{\statinst \in \protpcs} v(\statinst) \ge \targetval
\end{equation*}

FastFlip then efficiently selects the optimal $\protpcs$ for a range of
$\targetval$ values. This corresponds to solving the value / cost
multi-objective optimization problem using the $\epsilon$-constraint
method~\cite{Miettinen1998} (i.e., turning one of the objectives into a
constraint as we do above) to obtain Pareto-optimal choices for $\protpcs$.

\subsection{Precision of FastFlip}
\label{ff:sec:app:precision}

We observed that there are four major factors that affect the precision of FastFlip:

\paragraph{Inter-section masking.}
Inter-section masking occurs when an SDC in one section is masked by a
downstream section. FastFlip must conservatively assume that SDCs introduced in
any section result in SDCs in the final outputs. The frequency of inter-section
masking is highly dependent on the application.

\paragraph{Imprecision of component analyses.}
As FastFlip depends on the results of the error injection analysis and the SDC
propagation analysis, any imprecision in these analyses can lead to imprecision
in FastFlip. For example, in our evaluation, FastFlip is affected by:
\begin{itemize}
\item The error injection analysis's injection pruning heuristics, which make
per-instruction error injection practical at the cost of introducing some
inaccuracy in the injection outcomes~\cite[Figure~5]{approxilyzer}
\item The SDC propagation analysis's conservative SDC propagation, which causes
it to overestimate the magnitude of SDC at the end of the program
\end{itemize}

\paragraph{Side effects.}
FastFlip requires each analyzed program section to be free of side effects. If a
section modifies a variable that is used by a downstream section, then that
variable must be considered as an output of that section. Even after all such
variables are included in the output, the section may still cause additional
side effects as a result of errors. Below, we describe two major categories of
observed side effects that occur exclusively due to errors, along with the
strategies employed by FastFlip to account for them.

First, the error may cause the section to write to a memory location outside the
memory region where the section's output is stored (e.g., due to incorrect array
index calculation or evaluation of a loop exit condition). As a result, it is
possible for the outputs of the section to be correct while data in memory
locations adjacent to the outputs is corrupted. This data may be used by
downstream sections, leading to a side effect. FastFlip mitigates this effect by
considering variables stored in memory locations adjacent to the section outputs
to be an output of that section.

Second, the error may corrupt variables that will no longer be used by the
current section, but are used in downstream sections (e.g., due to the
corruption of a register being popped from the stack at the end of a section).
FastFlip mitigates this effect by including variables that will be used by
future sections in the output of the current section, even if the current
section only reads said variables. FastFlip uses the dataflow specification
provided by the developer to find such variables.

While these mitigation strategies eliminate a large majority of side effects
introduced as a result of errors, some such side effects can still occur. As a
result of these remaining side effects, the outcomes of some of the injections
recorded by FastFlip may be incorrect.

\paragraph{Untested error sites.}
A small number of error sites in the full program may not be included in any
program section. For example, if sections are executed multiple times within an
outer loop, then the instructions which increment the outer loop counter and
restart the outer loop body may be excluded from all program sections. FastFlip
conservatively assumes that, if an error occurs at such an excluded error site,
then it will \emph{always} produce an SDC-Bad outcome. More rigorously, FastFlip
creates an imaginary section $\trcsec_\bot$ containing all such untested error
sites $\aninj$ and assumes that $\forall \aninj \in \injset_{\trcsec_\bot}$,
$\outcome_\trcsec(\aninj) = \{ \infty, \ldots, \infty \}$. This reduces the
precision of FastFlip, as the actual number of SDC-Bad outcomes for these
untested error sites is often lower.

\subsection{Adapting FastFlip to compensate for loss of precision}
\label{ff:sec:app:vallosscomp}

If FastFlip loses precision as a result of the factors described in
Section~\ref{ff:sec:app:precision}, it can lead to a loss of \emph{utility}.
That is, a loss of precision can cause FastFlip to protect against a smaller
number of errors that cause SDC-Bad outcomes than expected. Similarly, it can
also increase the cost of protecting FastFlip's selection of static instructions
beyond the actual minimum cost of protection. FastFlip adaptively adjusts the
target value $\targetval$ used in Section~\ref{ff:sec:app:optimalprot} in order
to compensate for this loss of utility.

\paragraph{Measuring utility.}
FastFlip must first measure its loss of utility. FastFlip compares its utility
to the utility obtained via a baseline monolithic error injection analysis. The
baseline analysis injects errors in the whole program at once and directly
determines the outcome of these errors at the end of the program. It then uses
these results to selectively protect vulnerable static instructions. FastFlip
uses two primary metrics for measuring utility:

First, FastFlip treats the outcome labels of the monolithic error injection
analysis as the ground truth and calculates the value of protecting its
selection against SDC-Bad outcomes according to these alternate outcome labels.
FastFlip refers to the protection value of its selection calculated in this
manner as the \emph{achieved value} $\achvval$. FastFlip then compares
$\achvval$ to the target protection value $\targetval$. FastFlip calculates the
loss of value as $v_\textit{loss} = \targetval - \achvval$. Value loss measures
the degree to which FastFlip undershoots the target value of protection against
SDC-Bad outcomes; a lower value loss is better.

Second, FastFlip calculates the excess cost of FastFlip's selection over the
monolithic error injection analysis's selection. Specifically, if the costs
associated with protecting the two selections of instructions against SDCs are
$c_\textit{FF}$ (for FastFlip) and $c_\textit{Mono}$ (for the monolithic
analysis) respectively, the excess cost is $c_\textit{excess} = c_\textit{FF} -
c_\textit{Mono}$. Excess cost measures the degree of inefficiency of FastFlip's
selection for protecting against SDC-Bad outcomes as compared to the more
efficient selection made by the monolithic analysis; a lower excess cost is
better.

When analyzing a program, FastFlip can simultaneously run the monolithic error
injection analysis for minimal additional analysis time\footnote{For our
evaluation, we run the analyses separately for proper comparison of their
analysis time.}. To do so, FastFlip simultaneously checks the effect of each
error in each section both on the outputs of that section, as well as the final
outputs. Using these two sets of outcome labels, FastFlip can calculate
$v_\textit{loss}$ and $c_\textit{excess}$.

\paragraph{Adjusting the target value.}
FastFlip replaces the original target $\targetval$ with an adjusted target
$\targetval'$. Let the achieved value for this adjusted target be
$v'_\textit{achv}$. FastFlip minimizes $\targetval'$ such that $v'_\textit{achv}
\ge \targetval$. If $\targetval' > \targetval$, then the cost of protecting
FastFlip's selection increases, with larger adjustments leading to larger
increases. It is also possible that $\targetval' < \targetval$, in which case
the cost decreases instead.

\subsection{Composability and incremental analysis}

When developers modify a program section, FastFlip must rerun the error
injection and local sensitivity analysis on the modified program section and any
downstream sections whose input changes as a result of the modification.
FastFlip can reuse the results of these sub-analyses for all other sections.
Lastly, FastFlip must recalculate the end-to-end SDC propagation specifications
using the SDC propagation analysis. Since running the error injection analysis
is the major contributor to FastFlip's runtime, this approach leads to
significant speedups as compared to rerunning the error injection analysis on
the full modified program, even if FastFlip must re-analyze multiple sections
due to modifications that change the inputs of downstream sections.

\begin{algorithm}
  \begin{algorithmic}[1]
    \Input{$P_\textit{adj}$: target adjustment interval;
           $m_\textit{adj}$: number of modifications since last target adjustment;
           $\targetval$: original target value;
           $\targetval'$: adjusted target value}
    \Modifies{$m_\textit{adj}$; $\targetval'$}
    \Returns{$\protpcs$: selection of instructions to protect}
    \If{$m_\textit{adj} \ge P_\textit{adj}$}
      \State $m_\textit{adj} \gets 0$
      \State $\textit{Outcomes}_\textit{FF}, \textit{Outcomes}_\textit{Mono} \gets \Call{FastFlipAndMonolithic}{\textit{Program}, \textit{Input}}$
      \State $\targetval' \gets \Call{AdjustTarget}{\targetval, \textit{Outcomes}_\textit{FF}, \textit{Outcomes}_\textit{Mono}}$
    \Else
      \State $m_\textit{adj} \gets m_\textit{adj} + 1$
      \State $\textit{Outcomes}_\textit{FF} \gets \Call{FastFlipModifiedOnly}{\textit{Program}, \textit{Input}}$
    \EndIf
    \State $\protpcs \gets \Call{Knapsack}{\targetval', \textit{Outcomes}_\textit{FF}}$
  \end{algorithmic}
  \caption{Using adjusted target values when programs are modified}
  \label{ff:code:app:readjust}
\end{algorithm}

Algorithm~\ref{ff:code:app:readjust} shows how FastFlip uses target value
adjustment to compensate for loss of utility
(Section~\ref{ff:sec:app:vallosscomp}) when the program is modified. FastFlip
maintains a count of the number of modifications that have occurred since the
most recent target adjustment ($m_\textit{adj}$). If $m_\textit{adj}$ is below a
threshold chosen by the developer ($P_\textit{adj}$), FastFlip executes only its
own time saving compositional analysis and uses the existing adjusted target
$\targetval'$ to choose a set of static instructions to protect. That is, it is
not necessary to always run the monolithic analysis as described in
Section~\ref{ff:sec:app:vallosscomp} for modified programs. As developers make
modifications to the program, $\targetval'$ may no longer provide the expected
compensation for loss of utility. For this reason, once $m_\textit{adj} \ge
P_\textit{adj}$, FastFlip re-adjusts the target value by performing a fresh
analysis of the whole program while simultaneously running the monolithic
analysis. Developers can choose $P_\textit{adj}$ to trade off between utility
loss and analysis time.

\subsection{Characteristics of usable sub-analyses}
\label{ff:sec:app:usabletools}

\paragraph{Error injection analyses.}
The error injection analysis must separately report the outcome for errors in
each instruction in the program that the developer may wish to protect
(e.g.,~\cite{relyzer, approxilyzer}). Analyses that use sampling and only report
overall outcome statistics for the program (e.g.,~\cite{llfi, avgi}) are
incompatible with FastFlip, as they do not provide instruction-specific
vulnerability information. The analysis may inject single or multi-bit errors
into one dynamic instruction per simulation.

\paragraph{SDC propagation analyses.}
The SDC propagation analysis must support the same application-specific SDC
magnitude metric that the error injection analysis and the sensitivity analysis
use to report the SDC magnitude. The analysis must also support the propagation
of SDCs whose magnitude is represented by a symbolic variable, to enable the
calculation of Equation~\ref{ff:eq:app:erre2etotal}. Examples of supported
analyses are Chisel~\cite{chisel} and DeepJ~\cite{deepj}.

\section{Methodology}

\subsection{Choice of sub-analyses}
\label{ff:sec:met:tools}

\paragraph{Approxilyzer.}
Approxilyzer~\cite{approxilyzer} is an error injection analysis that focuses on
single bitflip errors that occur in CPU registers within in each dynamic
instruction in a program execution. It uses various heuristics to form
equivalence classes of bitflips that will most likely cause similar outcomes.
Approxilyzer injects a bitflip into the program execution for only one pilot
from each equivalence class. It then continues the now tainted program execution
(with possibly incorrect control flow), and records the outcome of the bitflip.
Finally, Approxilyzer applies the outcome of this pilot bitflip to all members
of the equivalence class. We specifically use gem5-Approxilyzer~\cite{gem5A}.

\paragraph{Chisel.}
Chisel~\cite{chisel} is an SDC propagation analysis that calculates the
end-to-end SDC propagation specification functions
$f_{\fulltrc,\lambda,\trcsec}$ as affine functions of the symbolic SDC
variables\footnote{We modified Chisel to add support for such symbolic SDC
variables.} $\errvar_{\trcsec,*}$ (Equation~\ref{ff:eq:app:erre2etotal}). Chisel
generates conservative specifications because it assumes that 1)~each program
section always amplifies input SDCs by the maximum amplification factor for that
section for any input, and 2)~when input SDCs can propagate through multiple
control flow paths, they are always affected by the maximum amplification factor
among all these paths, regardless of the actual path that the program execution
takes.

\subsection{Error model}
\label{ff:sec:met:errmdl}

As we compare FastFlip's results to those of an Approxilyzer-only approach, we
use the same error model as Approxilyzer (described below) to ensure a fair
comparison. We inject one single-bit transient error per simulation in an
architectural general purpose or SSE2 register. We target both source and
destination registers in dynamic instructions within the region of interest (a
subset of the correct dynamic trace of the program execution for a particular
input), and skip instructions without register operands. We do not inject errors
in special purpose, status, and control registers (e.g., \%rsp, \%rbp, and
\%rflags) as we assume that they always need protection which can be provided by
hardware. Similarly, we assume that caches are protected via hardware techniques
like ECC. As in previous works (e.g.,~\cite{vtrident, llfi}) we assume that the
probability $p(\aninj)$ that the error will occur at any error site $\aninj$
follows a uniform distribution.

\subsection{SDC detection model}

\paragraph{SDC detection mechanism.}
We assume that a compiler pass (e.g.,~\cite{swift, drift}) duplicates the
instructions that FastFlip selects for protection and then follows the original
and duplicate instructions with a check that ensures that their results match.
The duplicated code execution and increased register pressure leads to runtime
overhead. However, by allowing the compiler and CPU scheduler to rearrange
instructions and by coalescing multiple checks together, the overhead for
extensive instruction duplication across the program can be reduced to 29\% on
average~\cite{drift}. The overhead for selective duplication of individual
instructions is even lower.

\paragraph{Value and cost of SDC detection.}
We adapt the value and cost model from~\cite{HariAdve2012}, described below.
Since the error model in Section~\ref{ff:sec:met:errmdl} assumes that errors are
uniformly distributed ($p(\aninj)$ is uniform), the value of protecting a static
instruction $\statinst$ is proportional to the number of distinct errors
possible in $\statinst$ that produce an SDC-Bad outcome. The cost $c(\statinst)$
of protecting $\statinst$ is proportional to the number of dynamic instances of
$\statinst$ in the program execution trace. Developers can use alternate value
models by changing the error probability distribution $p(\aninj)$ and alternate
cost models by changing $c(\statinst)$.

\subsection{Benchmarks}

\begin{table}
  \centering
  \caption{List of benchmarks for FastFlip}
  \label{ff:tab:met:bench}
  \begin{tabular}{lrrr}
     \textbf{Benchmark} & \textbf{Input size} & \textbf{Sections} & $\mathbf{|\,\injset\,|}$ \\
     \midrule
     BScholes & 2 options       & 4 ($\times 2$) & 36.7K \\
     Campipe  & $32 \times 32$  & 5 ($\times 1$) & 72.7M \\
     FFT      & $256 \times 2$  & 5 ($\times 1$) & 9.23M \\
     LU       & $16 \times 16$  & 4 ($\times 2$) & 1.75M \\
     SHA2     & 32 bytes        & 3 ($\times 1$) &  403K \\
  \end{tabular}
\end{table}

Table~\ref{ff:tab:met:bench} shows our benchmarks for FastFlip. Column~2 shows
the input size, Column~3 shows the number of static sections (and the number of
dynamic instances of those sections), and Column~4 shows the total number of
error sites ($|\injset|$). We briefly describe the benchmarks and their origin
below:
\begin{itemize}
\item \emph{BScholes}: Black-Scholes stock option analysis benchmark from the
PARSEC suite~\cite{bienia2008parsec}
\item \emph{Campipe}: The raw image processing pipeline for the Nikon-D7000
camera, adapted from~\cite{cavapipe}
\item \emph{FFT}: Fast Fourier Transform benchmark from the Splash-3
suite~\cite{splash3}
\item \emph{LU}: Blocked matrix decomposition benchmark from the Splash-3
suite~\cite{splash3}
\item \emph{SHA2}: The SHA-256 hash function, adapted from~\cite{shaimpl}
\end{itemize}
For FFT and LU, the input size is the same as the minimized input size found by
Minotaur~\cite{minotaur}, a technique for reducing injection analysis time by
minimizing inputs without sacrificing program counter coverage. For BScholes, we
manually reduced the 21 option minimized input found by Minotaur down to 2
options while ensuring that the program counter coverage remained the same. For
Campipe, we use a reference $32 \times 32$ raw image input provided along with
the implementation. For SHA2, we use a common cryptographic key size (256 bits).

\subsection{Baseline, comparison, and experimental setup}
\label{ff:sec:met:compsetup}

\paragraph{Software and hardware.}
FastFlip uses gem5-Approxilyzer version 22.1~\cite{gem5A} simulating an x86-64
CPU as the architecture simulator. We performed our experiments on AMD Epyc
processors with 94 error injection experiment threads.

\paragraph{Region of interest.}
We focus on the computational portion of each benchmark and do not analyze I/O,
initial setup, or final cleanup code.

\paragraph{SDC magnitude metric.}
We use the maximum element-wise absolute difference as the SDC magnitude metric
for all benchmarks. Specifically, if $o_k[\ell]$ represents the
$\ell^\textit{th}$ element of an output $o_k$ and the modified output due to an
injection is $\hat{o}_k$, then the SDC metric is $\max_\ell |o_k[\ell] -
\hat{o}_k[\ell]|$.

\paragraph{SDC-Bad threshold.}
We first analyze all benchmarks assuming that any SDC is SDC-Bad, no matter how
small ($\forall \lambda.\ \errlim_\lambda = 0$). Next, we relax this requirement
by considering SDC magnitudes up to 0.01 to be tolerable, i.e., SDC-Good
($\forall \lambda.\ \errlim_\lambda = 0.01$) for all benchmarks except SHA2
(whose applications require the output to be fully precise).

\paragraph{Sensitivity analysis parameters.}
As we consider the maximum tolerable SDC magnitude $\errlim_\lambda$ to be 0.01,
we use this as the maximum perturbation during sensitivity analysis. To estimate
the Lipschitz constant $K$, we perform $10^6$ random perturbations up to
$\errlim_\lambda$. For array inputs, we randomly perturb one, multiple, or all
elements.

\paragraph{Comparison metrics.}
We compare the performance and utility of FastFlip to a baseline monolithic
Approxilyzer-only approach. The baseline approach treats the entire program as a
single section. For performance, we compare the analysis times of FastFlip and
Approxilyzer.

For comparing utility, we compare the selections of instructions to protect made
by the two approaches using the value loss and excess cost metrics described in
Section~\ref{ff:sec:app:vallosscomp}. FastFlip always uses target adjustment in
our evaluation, except in Section~\ref{ff:sec:eval:adjust}, where we investigate
the effects of target adjustment. We compare utility for four target values:
$\targetval \in \{0.90, 0.95, 0.99, 1.00\}$, which are target values that
correspond to protecting against 90\%, 95\%, 99\%, and 100\% of errors that
cause SDC-Bad outcomes, respectively.

\paragraph{Error range.}
While Approxilyzer's use of equivalence classes as described in
Section~\ref{ff:sec:met:tools} speeds up analysis, the pilot is not a perfect
predictor of the outcomes for the pruned injections (i.e., the rest of the
equivalence class). Figure~5 in Approxilyzer~\cite{approxilyzer} shows that, on
average, 4\% of pruned injections have an outcome that is significantly
different from the pilot. Therefore, Approxilyzer's results cannot be considered
to be the absolute ground truth for comparison.

To account for the potential discrepancy between the ground truth and
Approxilyzer, we calculate an error range around the value of SDC protection
according to Approxilyzer. Using the detailed results from Approxilyzer or
FastFlip, we can determine, for each error site, 1)~whether the outcome is
SDC-Bad, or a different outcome, and 2)~whether an injection was actually
performed at that error site, or the error site was pruned (the outcome for an
error at that site was inferred from the outcome for another error site in the
same equivalence class). Additionally, each error site can either be protected
or unprotected, depending on whether the analysis selected the corresponding
static instruction for protection.

Based on these classifying factors, we can divide all error sites into eight
categories, and use a variable to represent the \emph{number} of error sites in
each of these categories:\\
\begin{minipage}{\textwidth}
\centering
\vspace{2ex}
\begin{tabular}{lllll}
  & \multicolumn{2}{c}{\textbf{Injected}} & \multicolumn{2}{c}{\textbf{Pruned}} \\
  \cmidrule(lr){2-3} \cmidrule(lr){4-5}
  \textbf{Outcome:} & \textbf{SDC-Bad} & \textbf{Other} & \textbf{SDC-Bad} & \textbf{Other} \\
  \midrule
  \textbf{Protected} & $A$ & $B$ & $C$ & $D$ \\
  \textbf{Unprotected} & $E$ & $F$ & $G$ & $H$ \\
\end{tabular}
\vspace{2ex}
\end{minipage}
Lastly, let $R$ be the rate of outcome misprediction for pruned error sites.
Using the above information, we can calculate the lower and upper bounds on the
actual value of protecting the selected instructions:
\begin{align*}
v_{\min} &= \frac{A + (1-R)C}{A + (1-R)C + E + G + RH} \\
v_{\max} &= \frac{A + C + RD}{A + C + RD + E + (1-R)G}
\end{align*}
Then, the value error range is the interval $[v_{\min}, v_{\max}]$. Note that
using the above notation, the value calculated in
Section~\ref{ff:sec:app:optimalprot} is equal to:
\begin{equation*}
v_\textit{calc} = \frac{A + C}{A + C + E + G}
\end{equation*}
which always lies within $[v_{\min}, v_{\max}]$.

For FFT, LU, and BScholes, we use the benchmark-specific pilot prediction
inaccuracy from Figure~5 in Approxilyzer~\cite{approxilyzer} (3\%, 4\%, and 10\%
respectively). For Campipe and SHA2, we consider the average inaccuracy from the
same figure (4\%).

We calculate FastFlip's error range in this manner around its results for
achieved protection value. If, for this error range, $v_{\max} \ge \targetval$,
then we consider FastFlip's result to be acceptable, even if $\achvval$ is less
than $\targetval$.

\paragraph{Analysis time.}
For FastFlip, the total time required is the sum of the time required for
1)~analyzing each program section with Approxilyzer and the sensitivity
analysis, 2)~Chisel error propagation analysis, 3)~calculating the value of
protecting static instructions, and 4)~solving the knapsack problem. For
modified benchmarks, we do not include the time required to analyze the
unmodified sections in the first category. For Approxilyzer, the total time
required is the sum of the time required for 1)~analyzing the full program and
2)~solving the knapsack problem.

\paragraph{Timeouts.}
FastFlip assumes that if the error causes the runtime of a program section to
exceed $5 \times$ the nominal runtime, then the execution times out, which is a
detected outcome. We use the same timeout rule for Approxilyzer.

\subsection{Program modifications}
\label{ff:sec:met:mods}

To test the advantages offered by FastFlip's compositional analysis, we modify
one or more sections within each benchmark. Then, we re-analyze the full
modified program with Approxilyzer, and compare this to FastFlip's re-analysis
of only the modified sections followed by SDC propagation.

For each benchmark, we experiment with two types of semantics-preserving
modifications. \emph{Small} modifications represent simple modifications that
developers or compilers may make while optimizing and maintaining the program.
Such modifications of up to 15 lines of code form a majority of open-source
commits~\cite{typicalcommit}. For the \emph{large} modifications, we replace a
program section with a lookup hashtable. The hashtable stores key-value pairs
that map inputs of that section to corresponding outputs. If the modified
section finds the current input in this table, it returns the corresponding
output. Otherwise, it executes the original section code. Lastly, for the
BScholes benchmark, we also experiment with a modification that uses
coarse-grained code duplication to detect the presence of SDCs.

\paragraph{Details of \emph{small} modifications.}
For Campipe and FFT, we store an expression used in multiple locations within
the section into a variable to improve code readability. For LU, we introduce a
specialized version of a section that reduces the number of necessary bounds
checks when it detects that the matrix size is a multiple of the block size (as
is the case for our input). For BScholes, we eliminate a redundant floating
point operation that occurs in some cases in the cumulative normal distribution
function. This change very slightly changes the semantics of the section due to
floating point imprecision. For SHA2, we similarly eliminate a redundant shift
operation (without changing semantics or making the runtime input-dependent).

\section{Evaluation}

\subsection{Similarity of injection outcomes of FastFlip and Approxilyzer}
\label{ff:sec:eval:overallstats}

\begin{table}
  \centering
  \caption{Injection outcome counts for unmodified benchmarks}
  \label{ff:tab:eval:injresults}
  \begin{tabular}{lrrrr}
  & & \multicolumn{3}{c}{\textbf{Outcome counts}} \\
  \cmidrule(lr){3-5}
  \textbf{Benchmark} & \textbf{Approach} & \textbf{Detected} & \textbf{SDC} & \textbf{Masked} \\
  \midrule
  \multirow{3}{*}{BScholes} & Approxilyzer & 6.14K & 19.1K & 11.5K \\
                            &     FastFlip & 5.87K & 19.6K & 11.3K \\
                            \cmidrule(lr){2-5}
                            &         diff & -274 & +503 & -229 \\
  \midrule
  \multirow{3}{*}{Campipe}  & Approxilyzer & 29.0M & 11.1M & 32.6M \\
                            &     FastFlip & 29.8M & 18.2M & 24.8M \\
                            \cmidrule(lr){2-5}
                            &         diff & +749K & +7.09M & -7.84M \\
  \midrule
  \multirow{3}{*}{FFT}      & Approxilyzer & 3.53M & 4.22M & 1.48M \\
                            &     FastFlip & 3.53M & 4.40M & 1.31M \\
                            \cmidrule(lr){2-5}
                            &         diff & -7.61K & +178K & -170K \\
  \midrule
  \multirow{3}{*}{LU}       & Approxilyzer &  878K &  809K & 65.4K \\
                            &     FastFlip &  879K &  815K & 56.8K \\
                            \cmidrule(lr){2-5}
                            &         diff & +1.43K & +6.50K & -8.57K \\
  \midrule
  \multirow{3}{*}{SHA2}     & Approxilyzer & 65.2K &  318K & 19.9K \\
                            &     FastFlip & 64.5K &  319K & 19.8K \\
                            \cmidrule(lr){2-5}
                            &         diff & -698 & +798 & -100 \\
  \end{tabular}
\end{table}

Table~\ref{ff:tab:eval:injresults} shows how injection outcome statistics
compare between FastFlip and the baseline Approxilyzer approach for the
benchmarks without modifications. Column~2 shows the approach and Columns~3-5
show the number of detected, SDC, and masked outcomes, respectively. The counts
for the modified benchmarks have a similar trend.

The clearest difference in the outcome counts is the lower number of masked
outcomes and the higher number of SDC outcomes identified by FastFlip. This is
often the result of errors that cause an SDC in the output of a program section,
which is then masked by a downstream section. Because FastFlip checks for SDCs
the output of each section and conservatively assumes that SDCs in a section's
output always propagate to the final output, it cannot identify such
inter-section masking. This is especially visible for Campipe, whose last
section is a nearest-value lookup table that masks many SDCs from upstream
sections.

For LU and BScholes, FastFlip does not analyze a small number of error sites
that are related to the iteration of the outer loop containing the analyzed
program sections. As described in Section~\ref{ff:sec:app:precision}, FastFlip
conservatively assumes that errors at these untested error sites always lead to
SDC-Bad outcomes. As a result, FastFlip may select such untested instructions
for protection even if errors in them do not result in SDCs.

\subsection{Utility of FastFlip compared to that of Approxilyzer}
\label{ff:sec:eval:utility}

\begin{table}
  \centering \small
  \caption{Comparison of FastFlip and Approxilyzer utility when all SDCs are
  unacceptable (SDC-Bad). A \tcmark{} indicates that the achieved value is
  within the value error range of FastFlip.}
  \label{ff:tab:eval:results0prec}
  \begin{tabular}{llllllll}
    & & \multicolumn{2}{c}{$\mathbf{\targetvalbf=0.90}$} & \multicolumn{2}{c}{$\mathbf{\targetvalbf=0.95}$} & \multicolumn{2}{c}{$\mathbf{\targetvalbf=0.99}$} \\
    \cmidrule(lr){3-4} \cmidrule(lr){5-6} \cmidrule(lr){7-8}
    \textbf{Benchmark} & \textbf{Modif.} & \textbf{Value} & \textbf{Cost (diff)} & \textbf{Value} & \textbf{Cost (diff)}
      & \textbf{Value} & \textbf{Cost (diff)} \\
    \midrule
    \multirow{3}{*}{BScholes} & None  & 0.901\tcmark & 0.635 ($+0.000$) & 0.950\tcmark & 0.717 ($+0.000$) & 0.990\tcmark & 0.827 ($+0.000$) \\
                              & Small & 0.899\tcmark & 0.634 ($+0.003$) & 0.950\tcmark & 0.713 ($+0.000$) & 0.990\tcmark & 0.821 ($+0.000$) \\
                              & Large & 0.898\tcmark & 0.669 ($+0.000$) & 0.949\tcmark & 0.753 ($+0.000$) & 0.991\tcmark & 0.849 ($+0.000$) \\
    \midrule
    \multirow{3}{*}{Campipe}  & None  & 0.915\tcmark & 0.611 ($+0.038$) & 0.950\tcmark & 0.676 ($+0.017$) & 0.991\tcmark & 0.807 ($+0.024$) \\
                              & Small & 0.924\tcmark & 0.611 ($+0.060$) & 0.954\tcmark & 0.678 ($+0.030$) & 0.990\tcmark & 0.807 ($+0.034$) \\
                              & Large & 0.912\tcmark & 0.760 ($+0.068$) & 0.961\tcmark & 0.819 ($+0.043$) & 0.993\tcmark & 0.899 ($+0.015$) \\
    \midrule
    \multirow{3}{*}{FFT}      & None  & 0.900\tcmark & 0.544 ($+0.011$) & 0.950\tcmark & 0.629 ($+0.002$) & 0.990\tcmark & 0.780 ($+0.000$) \\
                              & Small & 0.904\tcmark & 0.542 ($+0.010$) & 0.950\tcmark & 0.629 ($+0.004$) & 0.990\tcmark & 0.781 ($+0.002$) \\
                              & Large & 0.900\tcmark & 0.492 ($+0.001$) & 0.950\tcmark & 0.586 ($-0.000$) & 0.987\tcmark & 0.716 ($-0.016$) \\
    \midrule
    \multirow{3}{*}{LU}       & None  & 0.900\tcmark & 0.603 ($+0.000$) & 0.950\tcmark & 0.694 ($+0.000$) & 0.990\tcmark & 0.873 ($+0.000$) \\
                              & Small & 0.901\tcmark & 0.606 ($+0.002$) & 0.951\tcmark & 0.698 ($+0.002$) & 0.990\tcmark & 0.875 ($+0.001$) \\
                              & Large & 0.902\tcmark & 0.560 ($+0.002$) & 0.951\tcmark & 0.640 ($+0.003$) & 0.990\tcmark & 0.826 ($-0.001$) \\
    \midrule
    \multirow{3}{*}{SHA2}     & None  & 0.900\tcmark & 0.666 ($+0.001$) & 0.950\tcmark & 0.772 ($+0.000$) & 0.990\tcmark & 0.908 ($+0.001$) \\
                              & Small & 0.900\tcmark & 0.665 ($+0.000$) & 0.949\tcmark & 0.771 ($-0.001$) & 0.990\tcmark & 0.908 ($+0.000$) \\
                              & Large & 0.883\tcmark & 0.476 ($-0.007$) & 0.943\tcmark & 0.551 ($-0.003$) & 0.985\tcmark & 0.655 ($-0.007$) \\
  \end{tabular}
\end{table}

Table~\ref{ff:tab:eval:results0prec} compares the utility of FastFlip and
Approxilyzer for selective protection of instructions against SDCs, using the
metrics described in Section~\ref{ff:sec:app:vallosscomp}. Columns~1-2 show the
benchmark name and version, respectively. Subsequent pairs of columns show the
utility comparison for the target protection values 0.90, 0.95, and 0.99,
respectively. In each column pair, the first column shows FastFlip's achieved
protection value. The second column shows the cost of protecting FastFlip's
selection, and compares this to the cost of protecting Approxilyzer's selection.

With target adjustment, FastFlip successfully meets all target values for the
original, unmodified version of each benchmark. Because FastFlip reuses the
adjusted targets for the modified versions as described in
Algorithm~\ref{ff:code:app:readjust}, it may not precisely meet the target for
those modified versions. The maximum difference between the target and achieved
values is 0.017 (1.7\%) for SHA2-Large. In all cases, the target value is within
FastFlip's achieved value error range (Section~\ref{ff:sec:met:compsetup}).

For most benchmarks, the cost of protecting FastFlip's selection of instructions
is at most 0.011 (1.1\%) more than the cost of protecting Approxilyzer's
selection. The exception is Campipe, for which FastFlip's cost can be higher by
as much as 0.068 (6.8\%). This is because, unlike the other benchmarks, FastFlip
has to aggressively adjust the target values for Campipe in order to compensate
for the loss of precision caused by inter-section masking and meet the original
target values. Section~\ref{ff:sec:eval:adjust} describes the consequences of
forgoing target adjustment.

We observed that if we removed the last section of Campipe (which is the primary
cause of high inter-section masking in Campipe), FastFlip's target adjustments
became less aggressive, and the excess cost of FastFlip decreased to 0.036
(3.6\%). This suggests that more precise SDC propagation analyses that also
calculate the probability of SDC masking may help to reduce the need for target
adjustment.

The geomean cost of protecting FastFlip's selection is 0.601, 0.685, and 0.819
for the target protection values 0.90, 0.95, and 0.99, respectively. This shows
that it is possible to protect against 90\% bitflips that cause SDCs by
protecting on average 60\% of all dynamic instructions. Protecting against the
remaining SDCs quickly leads to diminishing returns.

\subsection{Performance of FastFlip compared to that of Approxilyzer}

\begin{table}
  \centering
  \caption{Comparison of FastFlip and Approxilyzer analysis time}
  \label{ff:tab:eval:results0time}
  \begin{tabular}{llrrr}
    & & \multicolumn{3}{c}{\textbf{Analysis time (core-hours)}} \\
    \cmidrule(lr){3-5}
    \textbf{Benchmark} & \textbf{Modif.} & \textbf{FastFlip} & \textbf{Approxilyzer} & \textbf{Speedup} \\
    \midrule
    \multirow{3}{*}{BScholes} & None  &   69 hrs &   65 hrs & $0.9\times$ \\
                              & Small &   42 hrs &   62 hrs & $1.5\times$ \\
                              & Large &    3 hrs &   24 hrs & $8.4\times$ \\
    \midrule
    \multirow{3}{*}{Campipe}  & None  & 2459 hrs & 2631 hrs & $1.1\times$ \\
                              & Small &  158 hrs & 2720 hrs & $17.2\times$ \\
                              & Large &   45 hrs &  494 hrs & $11.0\times$ \\
    \midrule
    \multirow{3}{*}{FFT}      & None  &  980 hrs &  520 hrs & $0.5\times$ \\
                              & Small &  300 hrs &  509 hrs & $1.7\times$ \\
                              & Large &   93 hrs &  513 hrs & $5.5\times$ \\
    \midrule
    \multirow{3}{*}{LU}       & None  &  694 hrs &  602 hrs & $0.9\times$ \\
                              & Small &   80 hrs &  625 hrs & $7.8\times$ \\
                              & Large &   94 hrs &  441 hrs & $4.7\times$ \\
    \midrule
    \multirow{3}{*}{SHA2}     & None  &  726 hrs &  728 hrs & $1.003\times$ \\
                              & Small &  718 hrs &  726 hrs & $1.01\times$ \\
                              & Large &   43 hrs &   45 hrs & $1.05\times$ \\
  \end{tabular}
\end{table}

Table~\ref{ff:tab:eval:results0time} compares the analysis time of FastFlip
(without simultaneously running Approxilyzer) and Approxilyzer. Columns~1-2 show
the benchmark name and version, respectively. Columns~3-4 show the total
analysis time for FastFlip and Approxilyzer, respectively. Column~5 shows the
speedup of FastFlip over Approxilyzer. We measure analysis time in core-hours as
the error injection analysis is highly parallelizable. The actual analysis time
is much lower when using a large number of error injection experiment threads.

For FastFlip, over 99\% of analysis time is for the error injection analysis.
The sensitivity analysis requires less than five minutes. The symbolic SDC
propagation analysis and knapsack problem solver each require less than one
minute, even for programs and inputs that are much larger than our benchmarks.

The two approaches have similar analysis times for the unmodified (\emph{None})
versions of all benchmarks except FFT. For FFT, Approxilyzer prunes a larger
number of injections since operations such as transpose are performed multiple
times in different dynamic sections of the program execution. As FastFlip
injects errors into each section independently, it cannot similarly prune
injections across sections. However, FastFlip is faster when analyzing the
modified versions of FFT.

To enable target adjustment, FastFlip simultaneously runs the Approxilyzer
analysis as described in Section~\ref{ff:sec:app:vallosscomp}. We use the
methodology from~\cite[Section~4.7]{minotaur} to confirm that the time required
for this simultaneous approach is at most 1\% more than the greater of the
analysis times of FastFlip and Approxilyzer for the unmodified versions of the
benchmarks.

For the modified versions, the speedup depends on how much of the original
program the developer replaced. If the modified sections are small with respect
to the full program, then FastFlip must re-analyze only that small modified
portion of the program, as opposed to Approxilyzer which must re-analyze the
full modified program. This leads to the particularly large speedups for
Campipe. On the other hand, if the modified sections are a large portion of the
full program, then FastFlip must re-analyze large portions of the program,
leading to smaller speedups. This leads to the negligible speedups for SHA2,
where we modified the most expensive section of the program. As FastFlip reuses
the adjusted targets for modified benchmarks, it only needs to use the
simultaneous approach for the original version.

These results show that FastFlip can save a significant amount of analysis time
when analyzing modified programs. For modern software systems that accumulate
multiple small modifications over time, FastFlip can provide ever increasing
savings.

\subsection{Effects of target adjustment}
\label{ff:sec:eval:adjust}

\begin{table}
  \centering \small
  \caption{Comparison of FastFlip and Approxilyzer utility for Campipe with and
  without target adjustment. A \tcmark{} indicates that the achieved value is
  within the value error range of FastFlip, while a \txmark{} indicates the
  opposite.}
  \label{ff:tab:eval:resultswithcomp}
  \begin{tabular}{llllllllll}
    \textbf{Target type} & \textbf{Modif.} & \textbf{Value} & \textbf{Cost (diff)} & \textbf{Value} & \textbf{Cost (diff)}
      & \textbf{Value} & \textbf{Cost (diff)} \\
    \midrule
    & & \multicolumn{2}{c}{$\mathbf{\targetvalbf=0.90}$} & \multicolumn{2}{c}{$\mathbf{\targetvalbf=0.95}$} & \multicolumn{2}{c}{$\mathbf{\targetvalbf=0.99}$} \\
    \cmidrule(lr){3-4} \cmidrule(lr){5-6} \cmidrule(lr){7-8}
                      & None  & 0.848\txmark & 0.542 ($-0.031$) & 0.920\tcmark & 0.622 ($-0.038$) & 0.977\tcmark & 0.751 ($-0.032$) \\
    \textbf{Original} & Small & 0.879\tcmark & 0.543 ($-0.008$) & 0.925\tcmark & 0.623 ($-0.025$) & 0.980\tcmark & 0.752 ($-0.021$) \\
                      & Large & 0.868\txmark & 0.687 ($-0.005$) & 0.925\txmark & 0.776 ($+0.001$) & 0.979\tcmark & 0.864 ($-0.021$) \\
    \midrule
    & & \multicolumn{2}{c}{$\mathbf{\targetvalpbf=0.942}$} & \multicolumn{2}{c}{$\mathbf{\targetvalpbf=0.972}$} & \multicolumn{2}{c}{$\mathbf{\targetvalpbf=0.997}$} \\
    \cmidrule(lr){3-4} \cmidrule(lr){5-6} \cmidrule(lr){7-8}
                      & None  & 0.915\tcmark & 0.611 ($+0.038$) & 0.950\tcmark & 0.676 ($+0.017$) & 0.991\tcmark & 0.807 ($+0.024$) \\
    \textbf{Adjusted} & Small & 0.924\tcmark & 0.611 ($+0.060$) & 0.954\tcmark & 0.678 ($+0.030$) & 0.990\tcmark & 0.807 ($+0.034$) \\
                      & Large & 0.912\tcmark & 0.760 ($+0.068$) & 0.961\tcmark & 0.819 ($+0.043$) & 0.993\tcmark & 0.899 ($+0.015$) \\
  \end{tabular}
\end{table}

Section~\ref{ff:sec:eval:utility} presents the results of FastFlip when it uses
target adjustment in order to meet the original targets. For most benchmarks,
the adjusted targets are within 0.4\% of the original targets, so the results
with and without target adjustment are similar. For Campipe, FastFlip has to
aggressively adjust the target protection value because its characteristics lead
to precision loss.

Table~\ref{ff:tab:eval:resultswithcomp} compares the utility of FastFlip and
Approxilyzer for Campipe with and without target adjustment. The format is
similar to that of Table~\ref{ff:tab:eval:results0prec}, except that Column~1
indicates whether the results are for the original (non-adjusted) target or the
adjusted target. Without target adjustment, FastFlip undershoots the target by
as much as 0.052 (5.2\%) and the original target value does not always fall
within FastFlip's achieved value error range. These results demonstrate the
importance of target adjustment for ensuring that FastFlip meets the original
protection targets for all benchmarks. However, aggressive target adjustment
also leads to an increase in protection cost of FastFlip's selection of
instructions over Approxilyzer's selection.

\subsection{Comparison of FastFlip and Approxilyzer when small SDCs are acceptable}

\begin{table}
  \centering \small
  \caption{Comparison of FastFlip and Approxilyzer utility when SDCs greater
  than 0.01 are SDC-Bad. A \tcmark{} indicates that the achieved value is within
  the value error range of FastFlip.}
  \label{ff:tab:eval:results0.01}
  \begin{tabular}{llllllll}
    & & \multicolumn{2}{c}{$\mathbf{\targetvalbf=0.90}$} & \multicolumn{2}{c}{$\mathbf{\targetvalbf=0.95}$} & \multicolumn{2}{c}{$\mathbf{\targetvalbf=0.99}$} \\
    \cmidrule(lr){3-4} \cmidrule(lr){5-6} \cmidrule(lr){7-8}
    \textbf{Benchmark} & \textbf{Modif.} & \textbf{Value} & \textbf{Cost (diff)} & \textbf{Value} & \textbf{Cost (diff)}
      & \textbf{Value} & \textbf{Cost (diff)} \\
    \midrule
    \multirow{3}{*}{BScholes} & None  & 0.900\tcmark & 0.645 ($+0.013$) & 0.951\tcmark & 0.727 ($+0.013$) & 0.990\tcmark & 0.821 ($+0.000$) \\ 
                              & Small & 0.898\tcmark & 0.642 ($+0.011$) & 0.949\tcmark & 0.724 ($+0.013$) & 0.990\tcmark & 0.821 ($+0.000$) \\ 
                              & Large & 0.892\tcmark & 0.681 ($+0.012$) & 0.946\tcmark & 0.765 ($+0.006$) & 0.990\tcmark & 0.849 ($-0.012$) \\ 
    \midrule
    \multirow{3}{*}{Campipe}  & None  & 0.900\tcmark & 0.576 ($+0.031$) & 0.951\tcmark & 0.674 ($+0.032$) & 0.991\tcmark & 0.802 ($+0.030$) \\ 
                              & Small & 0.915\tcmark & 0.577 ($+0.057$) & 0.958\tcmark & 0.674 ($+0.054$) & 0.992\tcmark & 0.802 ($+0.045$) \\ 
                              & Large & 0.903\tcmark & 0.694 ($+0.024$) & 0.953\tcmark & 0.780 ($+0.018$) & 0.992\tcmark & 0.895 ($+0.016$) \\ 
    \midrule
    \multirow{3}{*}{FFT}      & None  & 0.900\tcmark & 0.563 ($+0.002$) & 0.950\tcmark & 0.687 ($+0.004$) & 0.990\tcmark & 0.848 ($+0.008$) \\ 
                              & Small & 0.904\tcmark & 0.579 ($+0.012$) & 0.947\tcmark & 0.684 ($-0.006$) & 0.989\tcmark & 0.845 ($+0.002$) \\ 
                              & Large & 0.895\tcmark & 0.529 ($-0.007$) & 0.936\tcmark & 0.625 ($-0.028$) & 0.979\tcmark & 0.774 ($-0.041$) \\ 
    \midrule
    \multirow{3}{*}{LU}       & None  & 0.900\tcmark & 0.657 ($+0.004$) & 0.950\tcmark & 0.787 ($+0.000$) & 0.990\tcmark & 0.932 ($+0.002$) \\ 
                              & Small & 0.906\tcmark & 0.674 ($+0.020$) & 0.950\tcmark & 0.787 ($+0.000$) & 0.990\tcmark & 0.932 ($+0.000$) \\ 
                              & Large & 0.903\tcmark & 0.638 ($+0.009$) & 0.943\tcmark & 0.753 ($-0.007$) & 0.989\tcmark & 0.884 ($-0.002$) \\ 
  \end{tabular}
\end{table}

Table~\ref{ff:tab:eval:results0.01} compares the utility of FastFlip and
Approxilyzer when small SDCs ($\le 0.01$) are considered acceptable (SDC-Good)
and the analyses focus on only protecting against larger SDCs (SDC-Bad).
Table~\ref{ff:tab:eval:results0.01} has the same format as
Table~\ref{ff:tab:eval:results0prec}. We exclude SHA2 for this evaluation as its
applications require the calculated hashes to be fully precise.

FastFlip successfully meets all target values for all benchmarks. The maximum
difference between the target and achieved values is 0.014 (1.4\%) for
FFT-Large. In all cases, the target value is within FastFlip's achieved value
error range caused by injection pruning. For most benchmarks, the cost of
protecting FastFlip's selection of instructions is at most 0.020 (2\%) more than
the cost of protecting Approxilyzer's selection. The exception is Campipe, for
which FastFlip's protection cost is higher by as much as 0.057 (5.7\%) due to
aggressive target adjustment.

The geomean cost of protecting FastFlip's selection is 0.619, 0.720, and 0.849
for the target protection values 0.90, 0.95, and 0.99, respectively. FastFlip
obtains the results in Table~\ref{ff:tab:eval:results0.01} at the same time as
the results in Table~\ref{ff:tab:eval:results0prec} for negligible additional
analysis time (less than one minute).

\subsection{Protecting against \emph{all} SDC outcomes}

\begin{table}
  \centering \small
  \caption{Comparison of FastFlip and Approxilyzer utility for protecting
  against \emph{all} SDC outcomes throughout the program execution. A \tcmark{}
  indicates that the achieved value is within the value error range of
  Approxilyzer.}
  \label{ff:tab:eval:resultsAllSDCs}
  \begin{tabular}{llllll}
    & & \multicolumn{2}{c}{$\mathbf{\targetvalbf=0.99}$} & \multicolumn{2}{c}{$\mathbf{\targetvalbf=1.00}$} \\
    \cmidrule(lr){3-4} \cmidrule(lr){5-6}
    \textbf{Benchmark} & \textbf{Modif.} & \textbf{Value} & \textbf{Cost (diff)} & \textbf{Value} & \textbf{Cost (diff)} \\
    \midrule
    \multirow{3}{*}{BScholes} & None  & 0.990\tcmark & 0.827 ($+0.000$) & 1.000\tcmark & 0.954 ($+0.010$) \\
                              & Small & 0.990\tcmark & 0.821 ($+0.000$) & 1.000\tcmark & 0.953 ($+0.011$) \\
                              & Large & 0.991\tcmark & 0.849 ($+0.000$) & 1.000\tcmark & 0.958 ($+0.024$) \\
    \midrule
    \multirow{3}{*}{Campipe}  & None  & 0.991\tcmark & 0.807 ($+0.024$) & 1.000\tcmark & 0.900 ($+0.002$) \\
                              & Small & 0.990\tcmark & 0.807 ($+0.034$) & 1.000\tcmark & 0.900 ($+0.018$) \\
                              & Large & 0.993\tcmark & 0.899 ($+0.015$) & 1.000\tcmark & 0.940 ($+0.000$) \\
    \midrule
    \multirow{3}{*}{FFT}      & None  & 0.990\tcmark & 0.780 ($+0.000$) & 1.000\tcmark & 0.971 ($+0.022$) \\
                              & Small & 0.990\tcmark & 0.781 ($+0.002$) & 1.000\tcmark & 0.971 ($+0.021$) \\
                              & Large & 0.987\tcmark & 0.716 ($-0.016$) & 1.000\tcmark & 0.932 ($+0.000$) \\
    \midrule
    \multirow{3}{*}{LU}       & None  & 0.990\tcmark & 0.873 ($+0.000$) & 1.000\tcmark & 0.982 ($+0.000$) \\
                              & Small & 0.990\tcmark & 0.875 ($+0.001$) & 1.000\tcmark & 0.983 ($+0.000$) \\
                              & Large & 0.990\tcmark & 0.826 ($-0.001$) & 1.000\tcmark & 0.932 ($+0.000$) \\
    \midrule
    \multirow{3}{*}{SHA2}     & None  & 0.990\tcmark & 0.908 ($+0.001$) & 1.000\tcmark & 0.989 ($+0.000$) \\
                              & Small & 0.990\tcmark & 0.908 ($+0.000$) & 1.000\tcmark & 0.989 ($+0.000$) \\
                              & Large & 0.985\tcmark & 0.655 ($-0.007$) & 1.000\tcmark & 0.723 ($+0.014$) \\
  \end{tabular}
\end{table}

Table~\ref{ff:tab:eval:resultsAllSDCs} compares the utility of FastFlip and
Approxilyzer for protecting against \emph{all} SDC outcomes. The format is
similar to that of Table~\ref{ff:tab:eval:results0prec}.
Table~\ref{ff:tab:eval:resultsAllSDCs} repeats the results for $\targetval=0.99$
for comparison.

FastFlip achieves a protection value of 1.000 (100\%) for $\targetval=1.00$. It
is not necessary (or possible) to adjust this target value. However, for some
benchmarks, FastFlip also selects for protection some instructions for which
Approxilyzer found only masked or detected outcomes. As a result, the overhead
of FastFlip's selection is still slightly higher than that of Approxilyzer's
selection. The geomean cost of protecting FastFlip's selection is 0.819 and
0.936 for the target protection values 0.99 and 1.00, respectively. There is a
significant increase in the number of dynamic instructions that must be
protected to protect against the last 1\% of SDC-causing errors. The results for
protecting against all SDC outcomes whose magnitude is greater than 0.01 are
similar.

\subsection{Case study: effectiveness of error detection mechanisms}
\label{ff:sec:eval:errdet}

To verify that error detection mechanisms can reduce the likelihood of SDC
outcomes as a result of errors, we modified one static section of the BScholes
benchmark. The modified section executes the original code twice and compares
the output of the two executions. If the outputs do not match, this indicates
that an error occurred, and the program raises an exception (a detected
outcome), as illustrated by the following pseudocode:\\
\begin{minipage}{\textwidth}
\begin{lstlisting}[numbers=none]
modifiedCode(input) {
  output1 = originalCode(input);
  output2 = originalCode(input);
  assert(output1 == output2);
  return output1;
}
\end{lstlisting}
\end{minipage}
We analyze the effects of errors in the modified section with FastFlip and
Approxilyzer to check the effectiveness of such an error detection mechanism.

\begin{table}
  \centering
  \caption{Effect of adding error detection mechanisms to a static section of BScholes}
  \label{ff:tab:eval:resultsErrDet}
  \begin{tabular}{lrrrr}
    & \multicolumn{1}{c}{\textbf{Error outcomes}} & \multicolumn{3}{c}{\textbf{Analysis time (core-hours)}} \\
    \cmidrule(lr){2-2} \cmidrule(lr){3-5}
    \textbf{Modif.} & \textbf{SDC / Total (\%)} & \textbf{FastFlip} & \textbf{Approxilyzer} & \textbf{Speedup} \\
    \midrule
    None            & 2991 / 5536 (54.0\%)            & 69 hrs & 65 hrs & $0.9\times$ \\
    Error detection &  882 / 8960 (\hphantom{0}9.8\%) & 21 hrs & 71 hrs & $3.4\times$ \\
  \end{tabular}
\end{table}

Table~\ref{ff:tab:eval:resultsErrDet} shows the results of adding this error
detection mechanism to BScholes. Column~2 shows the number of SDC outcomes as a
fraction of the total error outcomes. Columns~3-4 show the total analysis time
for FastFlip and Approxilyzer, respectively. Column~5 shows the speedup of
FastFlip over Approxilyzer. We find that the modification significantly reduces
the number and fraction of SDC outcomes occurring due to errors in the modified
section; this is despite the increase in code size due to duplication. The
remaining SDCs occur in 1)~non-duplicated instructions at the start or end of
the modified section, or 2)~instructions within the first execution of the
original code that cause side effects that affect the second execution too. As
with other modifications, FastFlip saves time over Approxilyzer because it must
only analyze the modified sections of the benchmark.

\section{Related work}

\paragraph{Error injection-less reliability analyses.}
ePVF~\cite{ePVF} is a dynamic analysis which finds locations where a bitflip
will cause a crash, as opposed to an SDC, with $\sim\!90$\% accuracy.
TRIDENT~\cite{trident} uses empirical observations of error propagation in
programs to predict the overall SDC probability of a program and the SDC
probabilities of individual instructions. Other works~\cite{sdc_tune, pvf} use
analytical modeling to detect SDCs in a program. Leto~\cite{boston2018leto} is
an SMT-based fault tolerance analysis that supports multiple error models,
including the single error model. However, it requires precise specifications of
program components, which are cumbersome for large programs when errors can
occur at any point. While these analyses can be faster than error injection
analyses, they are less accurate and may not be able to precisely estimate the
magnitude of the output SDC due to an error. FastFlip's compositional nature
makes error injection analysis more affordable by amortizing the cost of
analyzing evolving programs over time.

Aloe~\cite{aloe} statically analyzes programs with
potentially imperfect error recovery mechanisms to determine the overall
reliability of the program. Unlike FastFlip, Aloe supports error models where
multiple errors can simultaneously occur during program execution with varying
probabilities. This is a limitation of FastFlip, as the correctness of
Equation~\ref{ff:eq:app:errsectotal} relies on the assumption that only one
error occurs during program execution.

\paragraph{Error injection analyses.}
Error injection analyses operate at different levels of abstraction, including
hardware, assembly, and IR~\cite{ DweikRAEs, MeRLiN, swatsim, SmartInjector,
failstar, gefinmafin, gemfiParasyris, flipit, llfi, hamartia, avf_online, avgi}.
These analyses typically use \emph{sampling} - they select a statistically
significant number of error sites at random and only perform error injections at
those sites. While this is sufficient for providing overall outcome statistics
as we do in Section~\ref{ff:sec:eval:overallstats}, we cannot use such results
to determine which specific instructions or blocks of instructions are
particularly vulnerable to SDCs in order to protect them. However, FastFlip can
still use these analyses if they are modified to perform per-instruction error
injection like Approxilyzer~\cite{approxilyzer, gem5A}.

Minotaur~\cite{minotaur} reduces the size of inputs required to test the
reliability of programs when subjected to error injections, while keeping the
percentage of static instructions evaluated close to 100\% (as compared to the
reference input). We use the input sizes proposed by Minotaur for the FFT and LU
benchmarks, and manually minimize the Minotaur-proposed BScholes input to 2
options without sacrificing instruction coverage. While Minotaur reduces
injection analysis time by reducing input size, the program must still be
analyzed in full. FastFlip complements Minotaur by adding the flexibility of
only re-analyzing small sections of the program when developers modify it,
further reducing analysis time.

Papadimitriou and Gizopoulos~\cite{PapadimitriouDemystifying} show that
injecting errors in various SRAM hardware structures can give different results
compared to injecting errors at higher levels of abstraction. AVGI~\cite{avgi}
builds on~\cite{PapadimitriouDemystifying} to show that hardware errors manifest
in software in different ways, but result in similar distributions of final
outcomes across applications. Using this insight, AVGI accelerates
hardware-level fault injection for large workloads to provide overall outcome
statistics. Santos et al.~\cite{9505116} similarly examine how faults injected
at the RTL level affect common GPU instructions, and inject these
instruction-level effects into applications to provide overall outcome
statistics and identify vulnerable hardware components. However, FastFlip
requires the error injection analysis to report outcomes for errors at each
possible error site, as opposed to summary statistics
(Section~\ref{ff:sec:app:usabletools}). The above analysis techniques that aim
to efficiently determine the effect of low-level faults via hybrid fault
injection are too slow even for small program sizes when modified to report
outcomes in the manner required by FastFlip. If future analyses succeed in
providing such detailed outcome information for low-level faults in a scalable
manner, FastFlip would be able to use them to improve the accuracy of its
analysis.

\paragraph{SDC propagation analyses.}
SDC propagation analyses either propagate SDCs forward through
programs~\cite{diamont, uncertainT}, or propagate SDC bounds backwards through
programs~\cite{chisel, parallely, aloe}. While we instantiated FastFlip with the
Chisel~\cite{chisel} backwards SDC propagation analysis, the FastFlip approach
can instead use any alternate SDC propagation analysis that satisfies the
requirements outlined in Section~\ref{ff:sec:app:usabletools}.

Ashraf et al.~\cite{ashrafsc15} analyze the propagation of randomly injected
faults in MPI applications. Using the injection experiment results, they build a
model to estimate the number of memory locations corrupted over time to guide
roll-back decisions. Combining FastFlip with~\cite{ashrafsc15} is an attractive
opportunity for making per-instruction error injection analysis of MPI
applications practical.

Mutlu et al.~\cite{mutluhipc19} predict the effect of bitflips injected into
iterative applications on the final output by analyzing the effects of fault
injections on a limited number of iterations. While this potentially
gives~\cite{mutluhipc19} an advantage over FastFlip for applications that
iterate the same operation multiple times, unlike FastFlip,~\cite{mutluhipc19}
cannot handle applications with multiple sections that perform distinct
operations, such as our benchmarks.

\paragraph{Hardware-based selective protection.}
Researchers have examined the use of selective hardware hardening (e.g., via
redundancy or ECC) for improving hardware reliability while limiting the use of
additional chip area~\cite{9486703, 8554308, 5654485, 4556046}. These techniques
find and replicate only those hardware components that, as a result of transient
errors, produce unacceptable outcomes across the range of typical applications
that users are expected to run on the hardware. The error model we use for
FastFlip's evaluation (Section~\ref{ff:sec:met:errmdl}) also assumes that caches
and certain registers are protected within the hardware. FastFlip efficiently
provides information which can be used to apply \emph{additional},
software-based selective protection tailored to the needs of specific
applications, as opposed to adding further hardware protections irrelevant to
other applications.

\paragraph{Software-based selective SDC protection.}
Unlike crashes, timeouts, or clearly invalid data, SDCs are more difficult to
detect by nature. SWIFT~\cite{swift} uses instruction duplication to detect
errors in computational instructions. To reduce overhead, it makes use of
instruction reordering by the compiler and the processor. DRIFT~\cite{drift}
further reduces overhead by coalescing the checks of multiple duplicated
instructions, which reduces basic block fragmentation. SWIFT and DRIFT
\emph{completely} eliminate the possibility of SDCs occurring due to single
bitflip errors in the duplicated computational instructions. nZDC~\cite{nzdc}
provides comparable overhead to SWIFT while also protecting programs from 99.6\%
of SDCs caused by single bitflip errors during load, store, and control flow
instructions.

Shoestring~\cite{Shoestring2010} finds and duplicates only particularly
vulnerable instructions. Hari et al.~\cite{HariAdve2012} propose protecting
blocks of instructions with single detectors placed at the end of loops or
function calls. These two techniques use the results of error injection analyses
to guide \emph{selective} instruction duplication. We consider such techniques
to be \emph{clients} of FastFlip. They provide FastFlip with information about
the runtime overhead of protecting various instructions or instruction blocks.
Our evaluation in fact uses an adapted version of the value and cost model
from~\cite{HariAdve2012}. In return, FastFlip provides precise information on
which instructions should be protected in order to minimize runtime overhead
while protecting against a developer-defined fraction of SDC-causing errors.
After developers apply these techniques to protect FastFlip's selection of
instructions, FastFlip can re-analyze the protected sections to confirm the
decrease in SDC vulnerability. For FastFlip, we focused on efficiently handling
program modifications in general. We also briefly explore a modification that
reduces SDC vulnerability via coarse-grained code duplication in
Section~\ref{ff:sec:eval:errdet}. Analyzing sections after applying fine-grained
duplication of FastFlip's selection of vulnerable instructions is an interesting
topic for future work.

%% file: ff-defs.tex
\newcommand{\fulltrc}{T}
\newcommand{\trcsec}{s}

\newcommand{\injset}{J}
\newcommand{\aninj}{j}

\newcommand{\aninp}{i}
\newcommand{\anoutp}{o}

\newcommand{\pcmap}{\textit{PC}}
\newcommand{\statinst}{\textit{pc}}
\newcommand{\protpcs}{\textit{pc}_\textit{prot}}

\newcommand{\outcome}{\textit{O}}

\newcommand{\errvar}{\varphi}
\newcommand{\errlim}{\varepsilon}

\newcommand{\targetval}{v_\textit{trgt}}
\newcommand{\targetvalbf}{\mathbf{v_{trgt}}}
\newcommand{\targetvalpbf}{\mathbf{v'_{trgt}}}

\newcommand{\achvval}{v_\textit{achv}}